\documentclass[apj]{emulateapj}

\usepackage{graphicx}
\usepackage[varg]{txfonts}
\usepackage{float}
\usepackage[latin1]{inputenc}
\usepackage[normalem]{ulem}

\usepackage{natbib}
\newcommand{\FeH}{$\mbox{[Fe/H]}$} 
\newcommand{\AB}[2]{$\mbox{[#1/#2]}$} 

\newcommand{\FeHeq}[1]{$\mbox{[Fe/H]}={#1}$}    
\newcommand{\FeHsim}[1]{$\mbox{[Fe/H]}\sim{#1}$}  
\newcommand{\FeHlt}[1]{$\mbox{[Fe/H]}<{#1}$}    
\newcommand{\FeHle}[1]{$\mbox{[Fe/H]}\le{#1}$}   
\newcommand{\ABeq}[3]{$\mbox{[#1/#2]}={#3}$}    
\newcommand{\ABlt}[3]{$\mbox{[#1/#2]}<{#3}$}    
\newcommand{\ABgt}[3]{$\mbox{[#1/#2]}>{#3}$}    
\newcommand{\ABge}[3]{$\mbox{[#1/#2]}\ge{#3}$}   
\newcommand{\ABsim}[3]{$\mbox{[#1/#2]}\sim {#3}$} 
\newcommand{\ABlesssim}[3]{$\mbox{[#1/#2]}\lesssim {#3}$} 
\newcommand{\teffm}{T_{\mbox{\scriptsize eff}}}  

\newcommand{\logg}{\ensuremath{\log g}}

\newcommand{\mlp}{\ensuremath{\alpha_{\mathrm{MLT}}}}

\newcommand{\Tefft}{\ensuremath{T_{\mathrm{eff}}}}

\newcommand{\FAC}{HE~1327$-$2326} 
\newcommand{\CBB}{HE~0107$-$5240} 

\shortauthors{Li et al.}

\begin{document}

\title{Spectroscopic analysis of metal-poor stars from LAMOST: early results}

\author{Hai-Ning Li\altaffilmark{1}}

\author{Gang Zhao\altaffilmark{1}}

\author{Norbert Christlieb\altaffilmark{2}}

\author{Liang Wang\altaffilmark{1}}

\author{Wei Wang\altaffilmark{1}}

\author{Yong Zhang\altaffilmark{3}}

\author{Yonghui Hou\altaffilmark{3}}
       
       
\author{Hailong Yuan\altaffilmark{1}}

\altaffiltext{1}{Key Lab of Optical Astronomy, National Astronomical Observatories,
		Chinese Academy of Sciences, A20 Datun Road, Chaoyang, Beijing 100012, China}
\email{lhn@nao.cas.cn,gzhao@nao.cas.cn}
\altaffiltext{2}{Zentrum f{\"u}r Astronomie der Universit{\"a}t Heidelberg, Landessternwarte,
		K{\"o}nigstuhl 12, D-69117 Heidelberg, Germany}
\altaffiltext{3}{Nanjing Institute of Astronomical Optics \& Technology, National Astronomical Observatories, 
       Chinese Academy of Sciences, Nanjing 210042, China}
         
\received{ }
\accepted{ }

\begin{abstract}

We report on early results from a pilot program searching for metal-poor
stars with LAMOST and follow-up high-resolution observation acquired
with the MIKE spectrograph attached to the Magellan~II telescope.
We performed detailed abundance analysis for eight objects with iron abundances \FeHlt{-2.0},
including five extremely metal-poor (EMP; \FeHlt{-3.0}) stars
with two having \FeHlt{-3.5}.
Among these objects, three are newly discovered EMP stars,
one of which is confirmed for the first time with high-resolution spectral observations.
Three program stars are regarded as carbon-enhanced metal-poor (CEMP) stars,
including two stars with no enhancement in their neutron-capture elements, 
which thus possibly belong to the class of CEMP-no stars; one of these objects
also exhibits significant enhancement in nitrogen, and is thus a potential
carbon and nitrogen-enhanced metal-poor star. 
The \AB{X}{Fe} ratios of the sample stars generally agree with those
reported in the literature for other metal-poor stars in the same {\FeH}
range. We also compared the abundance patterns of individual program stars
with the average abundance pattern of metal-poor stars, and find only one
chemically peculiar object with abundances of at least two elements
(other than C and N) showing deviations larger than 0.5\,dex.
The distribution of \AB{Sr}{Ba} versus \AB{Ba}{H} agrees that
an additional nucleosynthesis mechanism is needed aside from a single $r$-process.
Two program stars with extremely low abundances of Sr and Ba
support the prospect that both main and weak $r$-process may
have operated during the early phase of Galactic chemical evolution.
The distribution of \AB{C}{N} shows that there are two groups of carbon-normal
giants with different degrees of mixing. However, it is difficult to
explain the observed behavior of the \AB{C}{N} of the nitrogen-enhanced
unevolved stars based on current data.

\end{abstract}


\section{INTRODUCTION}

Very metal-poor (\FeH \footnote{The notation of
 $[A/B]=\log(N_A/N_B)_{\star}-\log(N_A/N_B)_{\sun}$ is used here, where
 $N_A$ and $N_B$ are the number densities of elements A and B
 respectively,
and $\star$ and $\odot$ refer to the star and the Sun respectively}.
$<-2.0$) and extremely metal-poor (EMP; \FeHlt{-3.0}) stars
are believed to preserve in their atmospheres detailed information 
concerning the chemical compositions of the interstellar medium at the time and place
that they were born. These objects are expected to offer us a direct
view into the earliest phases of the evolution of the Galaxy, the formation
of the first generation of stars, and the earliest nucleosynthesis
events
\citep{McWilliam1995AJ,Norris1996ApJS,Beers&Christlieb2005ARAA,Frebel&Norris2013pss}.
Precise analysis of the chemical composition of metal-poor stars enables
us to indirectly probe the range of supernova nucleosynthesis yields
in the early Galaxy \citep[][and reference
 therein]{Heger&Woosley2010ApJ,Nomoto2013ARAA}, and to ultimately
constrain the primordial nucleosynthesis of cosmological models as well
\citep[for details, see][]{Bromm&Yoshida2011ARAA}.

Ever since the discovery of metal-poor stars by
\citet{Chamberlain&Aller1951ApJ}, numerous efforts have been devoted
to related fields and remarkable progress has been achieved. To explore
the nature of the first generation of stars, detailed abundance analyses
based on high-resolution spectroscopy have been performed to identify
various abundance patterns in metal-deficient stars
\citep[e.g.,][]{Ryan1996ApJ,Lai2008ApJ,Cohen2008ApJ,Norris2013ApJb}.
These explorations also resulted in the discovery of a variety of unusual
peculiarities in elemental abundances in metal-poor stars, such as large
enhancements in carbon and nitrogen relative to iron and the
abundance ratios seen in the Sun; enhancement of $\alpha$ elements; and
neutron-capture elements
\citep{Aoki2007ApJ,Pols2012AA,Aoki2013ApJ,Placco2013ApJ}. 
The ever-increasing sample size of metal-poor halo stars and 
the enrichment history of the Galactic halo have gradually been revealed 
through statistical investigations in the low-metallicity region of the metallicity
distribution function \citep[MDF;][]{Carollo2007Nature,Schoerck2009AA,Li2010AA,Yong2013ApJb}.

Recent studies using an increasing number of metal-poor stars
have made it clear that there exists a large fraction in such stars
with significant enhancements of carbon, i.e., carbon-enhanced
metal-poor (CEMP) stars. The frequency of CEMP stars is suggested
to increase with decreasing metallicity \citep{Cohen2005ApJL,
Lucatello2006ApJL,Carollo2012ApJ,Spite2013AA,Yong2013ApJb}.
Larger samples of CEMP stars also indicate different types of
chemical patterns, mainly including four sub-classes defined
based on the abundances of the neutron-capture elements
\citep{Beers&Christlieb2005ARAA}:
CEMP-s (enhanced in $s$-process elements),
CEMP-r (enhanced in $r$-process elements),
CEMP-rs (enhanced in both $s$- and $r$-process elements),
and CEMP-no (no enhancement in neutron-capture elements).
This diversity in chemistry suggests different sites of carbon production
in the early Galaxy, and detailed analyses of the elemental abundances
of CEMP stars allos us to understand the nature of their progenitor stars.

As a key to all these investigations, the number of metal-poor stars
has been tremendously increased by recent wide-angle sky surveys,
including the HK survey \citep{Beers1992AJ} and the Hamburg/ESO survey
\citep[HES;][]{Christlieb2008AA}, and more recently using data from the Sloan
Digital Sky Survey \citep{York2000AJ} and the Sloan Extension for
Galactic Understanding and Exploration \citep[SEGUE;][]{Yanny2009AJ,Rockosi2009astro}, 
as well as the SkyMapper Telescope
\citep{Keller2007PASA} using the novel photometric filter system.

Subsequently, high-resolution spectroscopic research has been performed
for the metal-poor stars found in these surveys \citep[e.g.,
][]{McWilliam1995AJ,Norris2001ApJ,Cohen2004ApJ,Barklem2005AA}.
Studies of large samples have been conducted, including the ``First
Stars'' project \citep{Cayrel2004AA,Bonifacio2009AA}, the ``0Z project''
\citep{Cohen2011rrls}, ``The Most Metal-Poor Stars'' project 
\citep{Norris2013ApJa,Yong2013ApJa}, and the ``Extremely Metal-Poor Stars
from SDSS/SEGUE'' \citep{Aoki2013ApJ} project. These efforts have
resulted in detailed investigations of the abundances of more than 200 EMP stars,
including two hyper metal-poor (HMP) stars \FAC ~(\FeHeq{-5.4},
\citealt{Frebel2005Nature}) and \CBB ~(\FeHeq{-5.3},
\citealt{Christlieb2002Nature}), and J0313$-$6708, a star at
\FeHlt{-7.1} which is suspected to have been pre-enriched by a single
supernova \citep{Keller2014Nature}.

These surveys and follow-ups have not only identified
chemically peculiar low-metallicity objects and made it possible to
explore the evolution of the Galactic halo in a much more detailed and
systematic manner, but have also led to a number of debates. Recent
discoveries of more EMP stars with \FeHlt{-3.5}
\citep{Yong2013ApJa,Placco2014ApJ} have enabled us to determine more
accurately the low-metallicity tail of the MDF of the Galactic halo. It
was found that the sharp cutoff at \FeHsim{-3.6} previously observed in
HES data \citep{Schoerck2009AA,Li2010AA} is probably an artifact caused
by small number statistics, and that the MDF in fact smoothly decreases
at least down to approximately \FeHsim{-4.1} \citep{Yong2013ApJb}. With
extended samples of EMP stars, \citet{Placco2014ApJ} also raises doubts
about the cutoff of the abundance ratio of \AB{Sr}{Ba} at about
\FeHsim{-3.6}, which was previously pointed out by
\citet{Aoki2013ApJ}. All of these unsolved questions urgently require 
additional EMP stars, especially objects with \FeHlt{-3.5}, to better
understand the truth and the real picture of the earliest phases of
Galactic chemical evolution.

The Large sky Area Multi-Object fiber Spectroscopic Telescope
(LAMOST, also known as the Wang-Su Reflecting Schmidt Telescope or the Guoshoujing Telescope)
\footnote{See \url{http://www.lamost.org/public/} for more detailed
 information, and the progress of the LAMOST surveys.}
is a new type of wide-field telescope with a large aperture \citep{Cui2012RAA}.
LAMOST started its pilot survey in 2010 \citep{Luo2012RAA},
and has conducted a five year regular survey since 2011. The
combination of a large aperture (4\,m) and high multiplex factor (4000
fibers) with a 5$\degr$ field of view makes it a unique facility. 
Its special design and high spectrum acquiring rate allow LAMOST to
efficiently carry out systematic spectroscopic surveys of the various
stellar components of the Galaxy \citep{Zhao2006ChJAA,Zhao2012RAA},
including metal-poor stars. Furthermore, with the low-resolution ($R =
1800$) spectroscopic data from LAMOST, it is possible to reliably
identify metal-poor stars in survey mode, thus evidently enhancing
the survey efficiency. Therefore, we are using this facility to further
increase the number of metal-poor stars.

This paper is organized as follows. Observations and data reduction are
addressed in Section~\ref{sec:obs}. The determination of stellar parameters
and abundance analysis are described in Section~\ref{sec:analysis}. 
The interpretations and discussions of the derived elemental abundances
including comparisons with literature values are presented in
Section~\ref{sec:results}. The conclusions are given in
Section~\ref{sec:conclusions}.

\section{TARGET SELECTION AND OBSERVATION}\label{sec:obs}

Sample stars were selected from the low-resolution database
($R=1800$) spectra acquired with LAMOST. After robust estimation of
stellar parameters and selection of candidates using LAMOST data,
high-resolution ($R\sim35,000$) spectroscopy was
obtained for eight interesting objects with Magellan/MIKE 
to carry out detailed abundance analysis of these stars.

\subsection{Low-resolution Spectroscopic Observation with LAMOST}\label{subsec:LRS-obs}

%
%

\begin{deluxetable*}{rccccrrr}
\tablenum{1}
\tablecolumns{8}
\tabletypesize{\scriptsize}
\tablewidth{0pc}
\tablecaption{Log of the Magellan/MIKE observations. \label{tab:MIKElog}}
\tablehead{
\colhead{ID}& \colhead{R.A.}& \colhead{Decl.}& \colhead{$r$}& \colhead{Date}& \colhead{Exptime}&
\colhead{S/N}& \colhead{$v_{r}$}\\
\colhead{}& \colhead{}& \colhead{}& \colhead{(mag)}& \colhead{}& \colhead{(s)}&
\colhead{(pixel$^{-1})$}& \colhead{(km s$^{-1}$)}}
\startdata
LAMOST J0006$+$1057&00 06 17.20&$+$10 57 41.8&12.59&2013 Dec&2700&86 &$-$281.6\\
LAMOST J0102$+$0428&01 02 12.66&$+$04 28 24.2&10.87&2013 Aug& 360&72 &$-$201.3\\
LAMOST J0126$+$0135&01 26 58.58&$+$01 35 15.4&12.30&2013 Dec&1800&75 &$-$197.3\\
LAMOST J0257$-$0022&02 57 06.93&$-$00 22 33.7&14.97&2013 Aug&1890&48 &     2.7\\
LAMOST J0326$+$0202&03 26 53.88&$+$02 02 28.1&11.55&2013 Dec&1200&106&   121.2\\
LAMOST J0343$-$0227&03 43 22.87&$-$02 27 56.1&13.43&2013 Dec&2400&62 & $-$ 4.3\\
LAMOST J1626$+$1721&16 26 14.78&$+$17 21 12.1&14.29&2013 Aug&1300&56 &$-$177.1\\
LAMOST J1709$+$1616&17 09 59.79&$+$16 16 13.3&13.13&2013 Aug& 900&43 &$-$324.2\\
\enddata
\tablecomments{{\footnotesize The S/N ratio per pixel was measured at $\lambda \sim 4500$\,{\AA}.}}
\end{deluxetable*}

%
%

\begin{deluxetable*}{lccrrcrcrcrcrcrcrcrcrcr}
\tablenum{2}
\tablecolumns{30}
\tabletypesize{\scriptsize}
\tablewidth{0pc}
\tablecaption{Equivalent width Measurements and Line-by-line Abundances.\label{tab:EW}}
\tablehead{
\colhead{}& \colhead{}& \colhead{}& \colhead{}&
\multicolumn{3}{c}{J0006$+$1057}& \colhead{}& \multicolumn{3}{c}{J0102$+$0428}& \colhead{}&
\multicolumn{3}{c}{J0126$+$0135}& \colhead{}& \multicolumn{3}{c}{J0257$-$0022}& \colhead{} \\
\cline{5-7}\cline{9-11}\cline{13-15}\cline{17-19}\\
\colhead{Ion}& \colhead{$\lambda$}& \colhead{$\chi$}& \colhead{log $gf$}&
\colhead{$W$}& \colhead{log$\epsilon$(X)}& \colhead{$\sigma$}& \colhead{}&
\colhead{$W$}& \colhead{log$\epsilon$(X)}& \colhead{$\sigma$}& \colhead{}&
\colhead{$W$}& \colhead{log$\epsilon$(X)}& \colhead{$\sigma$}& \colhead{}&
\colhead{$W$}& \colhead{log$\epsilon$(X)}& \colhead{$\sigma$} \\
\colhead{}& \colhead{(\AA)} &\colhead{(eV)}& \colhead{}&
\colhead{(m\AA)}& \colhead{}& \colhead{}& \colhead{}&
\colhead{(m\AA)}& \colhead{}& \colhead{}& \colhead{}&
\colhead{(m\AA)}& \colhead{}& \colhead{}& \colhead{}&
\colhead{(m\AA)}& \colhead{}& \colhead{}}
\startdata
\ion{Na}{1}&5889.95&0.00&   0.108&134.7&3.05&   0.09& &186.1&3.69&   0.09& &146.3&2.86&   0.11& & ... & ...&   ... \\
\ion{Na}{1}&5895.92&0.00&$-$0.194& ... &... &   ... & &155.8&3.50&$-$0.09& &112.7&2.65&$-$0.11& & ... & ...&   ... \\
\ion{Mg}{1}&3829.35&2.71&$-$0.208& ... &... &   ... & &198.4&5.36&   0.12& &134.0&4.38&$-$0.07& &119.7&5.30&$-$0.17\\
\ion{Mg}{1}&4167.27&4.35&$-$0.710& ... &... &   ... & & 52.8&5.15&$-$0.09& & ... & ...&   ... & & ... & ...&   ... \\
\ion{Mg}{1}&4702.99&4.33&$-$0.380& 61.5&4.95&$-$0.05& & 69.8&5.03&$-$0.22& & 26.7&4.26&$-$0.19& & 45.1&5.49&   0.02\\
\ion{Mg}{1}&5172.68&2.71&$-$0.450& ... &... &   ... & &211.2&5.36&   0.12& &161.7&4.49&   0.04& &117.1&5.40&$-$0.07\\
\ion{Mg}{1}&5183.60&2.72&$-$0.239&187.8&5.10&   0.09& &230.4&5.37&   0.13& &178.9&4.56&   0.11& & syn &5.55&   0.08\\
\ion{Mg}{1}&5528.40&4.34&$-$0.498& ... &... &   ... & & 80.4&5.25&   0.01& & ... & ...&   ... & & 43.5&5.59&   0.12\\
\ion{Al}{1}&3943.99&0.00&$-$0.640& ... &... &   ... & &154.4&3.56&$-$0.08& & syn &2.27&   0.08& & 60.6&3.43&$-$0.02\\
\ion{Al}{1}&3961.52&0.01&$-$0.340&112.5&2.61&   0.05& &182.4&3.72&   0.08& & syn &2.12&$-$0.08& & 77.3&3.46&   0.02\\
\ion{Si}{1}&3905.52&1.91&$-$1.092& ... &... &   ... & &203.1&5.22&   0.09& & ... & ...&   ... & &127.2&5.80&   0.07\\
\ion{Si}{1}&4102.94&1.91&$-$3.140& 61.9&4.72&   0.03& & 85.3&5.05&$-$0.09& & 61.9&4.47&   0.03& & ... & ...&   ... \\
\enddata
\tablecomments{{\footnotesize The three columns for each objects refer to the measured equivalent width,
corresponding abundance in log$\epsilon$(X), and the uncertainty caused by the equivalent width measurement.
The ``syn'' indicates that the abundance has been derived using spectral synthesis. \\
(This table is available in its entirety in a machine-readable form.)}}
\end{deluxetable*}

Our study is based on the first data release of the LAMOST survey 
which was internally released after the first year of LAMOST survey operations.
The wavelength coverage ($3700$--$9100$\,{\AA}) and resolving power ($R=1800$)
of the LAMOST spectra allow a robust estimation of the stellar parameters, including metallicities.
Three independent methods are used to determine the metallicity of an object.

The first method is an application of an updated version of the methods
described by \citet{Beers1999AJ}, which obtain {\FeH} by making use of
the \ion{Ca}{2}~K line index KP measuring the strength of this
line and the HP2 index measuring the strength of the H$_{\delta}$ line.
This method has also been adopted by previous investigations
such as \citet{Schoerck2009AA} to which we refer the reader for details.

The other two methods make use of a grid of synthetic spectra. These
spectra were computed by using the synthesis code
\texttt{SPECTRUM}~\citep{Gray&Corbally1994AJ}, starting from the ATLAS9
grid of stellar atmospheric models of \citet{Castelli&Kurucz2003IAUS} with 
linear interpolation in three dimensions, i.e., {\Tefft}, {\logg}, and
{\FeH}. A microturbulence velocity of 2\,km\,s$^{-1}$ was adopted for
all of the spectra. The line list and atomic data were taken from
\texttt{SPECTRUM}~\footnote{http://www.appstate.edu/~grayro/spectrum/spectrum.html}.
The synthetic spectra have a resolution of 0.01\,{\AA}, covering
$4000\,\mathrm{K}\le\teffm\le 9000\,\mathrm{K}$ in steps of 250\,K,
$0.0\le\logg\le 5.0$ in steps of 0.5\,dex, and
$-4.0\le\mathrm{[Fe/H]}\le-0.5$ in steps of 0.5\,dex. Considering the
fact that we are searching for candidate metal-poor stars, the
\AB{$\alpha$}{Fe} was set to $+0.4$\,dex, which is the typical value in this \FeH ~region
(e.g., \citealt{McWilliam1997ARAA}).

The second method uses comparisons of line indices.
Based on the most up-to-date wavelength definitions of Lick indices
\footnote{http://astro.wsu.edu/worthey/html/index.table.html},
we have calculated corresponding indices for all of the synthetic spectra,
resulting in a line index grid covering the corresponding space of stellar parameters.
The same series of indices are then calculated
for each observed spectrum, and compared with this synthetic grid
to find the best match of parameters.

The third method that we adopt is based on a direct comparison
of the normalized observed flux with the normalized synthetic spectra
in the wavelength range $4500\,\mathrm{\AA}\le\lambda\le 5500\,\mathrm{\AA}$.
The $\chi^{2}$ minimization technique of \citet{Lee2008AJ_SSPP1}
is used to find the best-matching set of parameters.

Taking into account the typical uncertainty of $\sim 0.1$ -- $0.3$\,dex
for metallicities derived from low-resolution spectra,
an object is regarded as an EMP candidate if it is within the temperature
range $4000\,\mathrm{K}<\teffm< 7000\,\mathrm{K}$, and if at least two of
the three methods described above yield \FeHle{-2.7}.
The constraint on {\Tefft} is adopted to exclude low-luminosity,
foreground late-type stars belonging to the
disk population(s), and to ensure that only stars redward of the
main-sequence turnoff for stars with ages of $\sim 13$\,Gyr
(as predicted by theoretical isochrones) enter the sample.
The spectra of the above automatically selected candidates were
visually inspected to remove false positives such as cool white dwarfs,
or objects that were selected because their spectra were disturbed by
reduction artifacts.

\subsection{High-resolution Spectroscopy with Magellan/MIKE}\label{subsec:HRS-obs}

For eight of candidates selected as described in
Section~\ref{subsec:LRS-obs}, high-resolution spectra were acquired
during two runs in 2013 August and 2013 December, with the Magellan
Inamori Kyocera Echelle spectrograph (MIKE; \citealt{Bernstein2003SPIE})
at the Magellan-Clay Telescope at Las Campanas Observatory.
Among these objects, J0126$+$0135 and J0326$+$0202 were previously
selected from HES \citep{Frebel2006ApJ}, also known as HE~0124$-$0119
and HE~0324$+$0152, respectively. For J0326$+$0202, high-resolution observation
and abundance analysis have been performed as described in \citet{Hollek2011ApJ},
and we include this object for comparison purpose.
For the latter object, J0126$+$0135,
since there has been no high-resolution observation to confirm this object,
we have also included it into our target list.

The setup of the observations included a 0\".7 slit with $2\times 2$ and $1\times
2$ binning CCD readout modes (used for the two runs respectively),
yielding resolving powers of $R\sim $35,000 in the blue
spectral range (3300 --- 4900\,\AA) and $R\sim$28,000 in the red spectral range (4900 --- 9400\,\AA).
The spectra have an average signal-to-noise ratio per pixel of $S/N \sim 75$ at
4500\,{\AA}. More information about the observations, as well as the target
coordinates and magnitudes, are listed in Table~\ref{tab:MIKElog}.

%
%

\begin{deluxetable*}{rcccccccc}
\tablenum{3}
\tablecolumns{9}
\tabletypesize{\scriptsize}
\tablewidth{0pc}
\tablecaption{Adopted Stellar Parameters of the Program Stars.\label{tab:stellar-param}}
\tablehead{
\colhead{}&\multicolumn{4}{c}{MIKE Measurements}&\colhead{}&\multicolumn{3}{c}{LAMOST Measurement}\\
\cline{2-5}\cline{7-9}\\
\colhead{ID}& \colhead{\Tefft}& \colhead{log $g$}& \colhead{\FeH}& \colhead{$\xi$} &
\colhead{}& \colhead{\Tefft}& \colhead{log $g$}& \colhead{\FeH}\\
\colhead{}& \colhead{(K)}& \colhead{}& \colhead{}& \colhead{(km $^{-1}$)} &
\colhead{}& \colhead{(K)}& \colhead{}& \colhead{}}
\startdata
LAMOST J0006$+$1057 & 4520 & 0.6 & $-3.26$ & 2.2 & & 4800 & 1.6 & $-3.03$\\
LAMOST J0102$+$0428 & 4540 & 0.6 & $-2.75$ & 2.5 & & 4940 & 1.2 & $-2.78$\\
LAMOST J0126$+$0135 & 4330 & 0.1 & $-3.57$ & 2.8 & & 4810 & 1.2 & $-3.15$\\
LAMOST J0257$-$0022 & 6330 & 4.2 & $-2.24$ & 1.5 & & 6260 & 3.6 & $-2.65$\\
LAMOST J0326$+$0202 & 4700 & 1.1 & $-3.36$ & 1.8 & & 4830 & 1.8 & $-3.15$\\
LAMOST J0343$-$0227 & 4790 & 1.5 & $-2.42$ & 2.0 & & 4795 & 1.8 & $-2.57$\\
LAMOST J1626$+$1721 & 5930 & 3.6 & $-3.20$ & 1.6 & & 5830 & 3.4 & $-3.00$\\
LAMOST J1709$+$1616 & 5780 & 3.5 & $-3.71$ & 1.9 & & 6070 & 3.0 & $-3.33$\\
\enddata
\end{deluxetable*}

The raw data were reduced with the standard \texttt{IRAF}
\footnote{IRAF is distributed by the National Optical Astronomy
 Observatory, which is operated by the Association of Universities for
 Research in Astronomy, Inc., under cooperative agreement with the
 National Science Foundation.} procedures to obtain 1D spectra that
are flat-fielded, wavelength-calibrated, co-added, and
continuum-normalized. The radial velocities were obtained using the
\texttt{IRAF} procedure \texttt{fxcor}, employing a synthetic spectrum
with low-metallicity as a template.

\begin{figure}
\epsscale{1.2}
\plotone{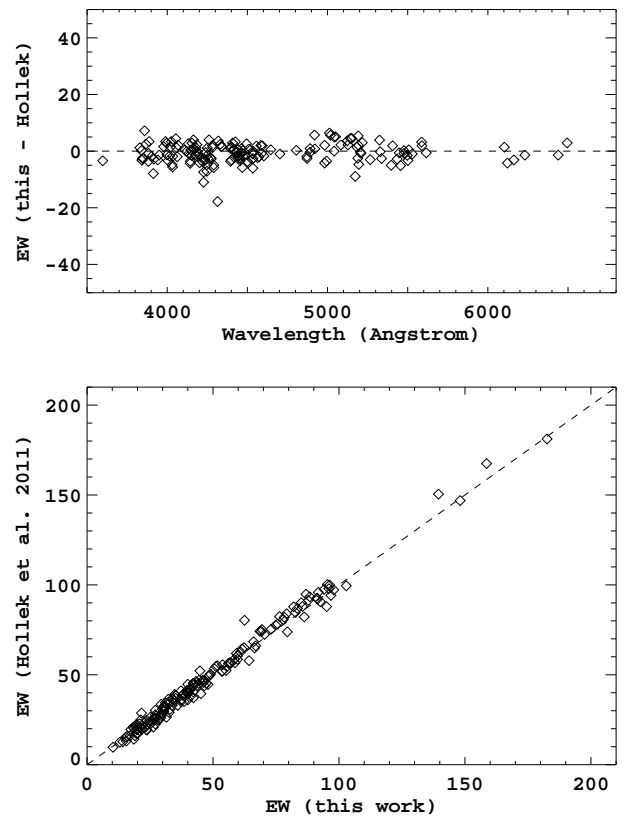}
\caption{Comparison of the equivalent widths of J0326$+$0202, respectively, measured by
this work and \citet{Hollek2011ApJ}.
Top: residuals of the equivalent widths (this work - Hollek) vs. the wavelength.
Bottom: direct comparison of the two measurements of the equivalent widths.
The one-to-one line is plotted for reference.
\label{fig:EW_compare}}
\end{figure}

Equivalent widths were measured by fitting Gaussian profiles to isolated
atomic absorption lines based on the line list of \citet{Frebel2013ApJ},
supplemented with \ion{K}{1}, \ion{Mn}{1}, and \ion{Y}{2} lines from
\citet{Placco2014ApJ}. Table~\ref{tab:EW} shows the measurements of the
adopted atomic lines for all of the program stars.
Figure~\ref{fig:EW_compare} compares our measured equivalent widths of J0326$+$0202
with those measured by \citet{Hollek2011ApJ}. There is a mean difference of
$-$0.6\,m{\AA} with $\sigma=3.4$\,m{\AA} between our measurements 
and those of \citet{Hollek2011ApJ} for lines included in our analysis.

\section{STELLAR PARAMETERS AND ABUNDANCE ANALYSIS}\label{sec:analysis}

For our abundance analysis, we use 1D plane-parallel, hydrostatic model
atmospheres of the ATLAS NEWODF grid of \citet{Castelli&Kurucz2003IAUS},
assuming a mixing-length parameter of $\mlp=1.25$, no convective
overshooting, and local thermodynamic equilibrium. We use an updated
version of the abundance analysis code MOOG \citep{Sneden1973ApJ}, which
treats continuous scattering as a source
function which sums both absorption and scattering
rather than true absorption \citep{Sobeck2011AJ}.

We adopt the photospheric Solar abundances of \citet{Asplund2009ARAA} when
calculating the [X/H] and [X/Fe] abundance ratios.

\subsection{Atmospheric Parameters}\label{subsec:atmos-param}

The effective temperatures {\Tefft} of the stars were determined by
minimizing the trend of the relationship between the derived abundances
and the excitation potentials of \ion{Fe}{1} lines. Previous investigations
and experience have shown that this procedure yields effective
temperatures with systematic offsets compared to those determined
from e.g., broadband optical and infrared photometry. Based on a
literature sample of metal-poor stars, \citet{Frebel2013ApJ} derives a
linear relation between the spectroscopic and photometric effective
temperatures, and this relation can be used to correct such systematic
deviations. Therefore, we have adopted this correction to obtain the
final effective temperatures of our sample.

Microturbulent velocities $\xi$ were determined from the analysis of
\ion{Fe}{1} lines, by forcing the iron abundances of individual lines to
exhibit no dependence on the reduced equivalent widths. For objects
that have a sufficient number of \ion{Fe}{2} lines detected, the
surface gravity {\logg} was determined by minimizing the difference
between the average abundances derived from the \ion{Fe}{1} and \ion{Fe}{2}
lines.

%
%

However, there are two objects at low metallicities, namely J1626$+$1721 and J1709$+$1616,
for which only spectra of lower-than-average S/N
are available; therefore only one or no \ion{Fe}{2} lines
could be detected in their spectra.
Therefore, we resorted to the theoretical isochrones of
\citet{Demarque2004ApJS} to estimate the surface gravities for these
objects given their temperatures and metallicities.
Isochrones with an age of 13\,Gyr were adopted, and all three
parameters (i.e., {\Tefft}, {\logg}, $\xi$) were iterated until
convergence was reached, i.e., with the derived surface gravity,
the ``zero'' slope can be obtained for derived abundances versus excitation potentials,
and abundances versus EWs, for corresponding {\Tefft} and $\xi$.
For these two objects, two solutions of surface gravity could be deduced
from the theoretical isochrones for each set of temperature and metallicity,
e.g., {\logg} of 4.57 and 3.62 for J1626$+$1721,
and 4.63 and 3.52 for J1709$+$1616. However, only the sub-giant {\logg} value
could result in convergence during the iterations, and therefore 3.62 and 3.52
were adopted for J1626$+$1721 and J1709$+$1616, respectively.

%
%
\begin{figure}[htbp]
\epsscale{1.2}
\plotone{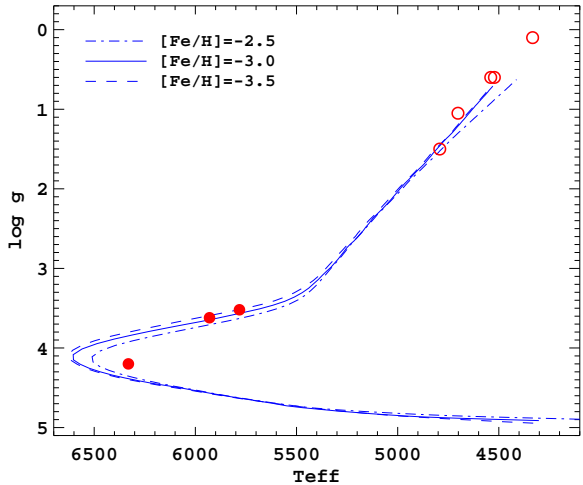}
\caption{\Tefft ~vs. \logg ~of the sample.
Filled circles correspond to carbon-enhanced objects, and the open circles refer to the carbon-normal ones,
whose definition is described in Section~\ref{subsec:CandN}.
For reference, the 13\,Gyr theoretical isochrones with \ABeq{$\alpha$}{Fe}{+0.4} and \FeHeq{-2.5}, $-3.0$,
and $-3.5$ from \citet{Demarque2004ApJS} are plotted with dash-dotted, solid and dashed lines respectively. 
\label{fig:HR}}
\end{figure}

During the iterations for determining the stellar parameters,
any Fe line with an offset of more than $3\sigma$ from the mean abundance was rejected.
The adopted stellar parameters of all the program stars are listed in
Table~\ref{tab:stellar-param}.
The derived stellar parameters
of J0326$+$0202 are very close to the values given by \citet{Hollek2011ApJ},
with a difference of $-75$\,K in {\Tefft}, $-0.1$ in {\logg}, and $-0.1$ in {\FeH}, respectively.
The parameters derived from LAMOST low-resolution spectra are also shown,
using the average values for the parameters derived from the three methods
as described in Section~\ref{subsec:LRS-obs}.
A relatively large discrepancy is found in {\logg}
for the cool giants, which is partly due to the large steps (0.5\,dex)
in {\logg} of the adopted synthetic spectra template.
The general agreement between the two sets of parameters shows that the LAMOST
selection pipeline is able to derive reliable parameters, especially
{\Tefft} and {\FeH}. In addition, the {\FeH} derived from LAMOST spectra
for J0126$+$0135 ($-3.15$) is quite consistent with that from HES follow-up observations 
($-3.1$, \citealt{Frebel2006ApJ}).

The distribution of {\Tefft} versus {\logg}
of all of the program stars is plotted in Figure~\ref{fig:HR},
with filled and open circles referring to the carbon-enhanced and carbon-normal
stars in the sample (as will be described in Section~\ref{subsec:CandN}); 
the theoretical isochrones from \citet{Demarque2004ApJS} are also shown for reference.

%
%
\begin{figure*}
\epsscale{1.0}
\plotone{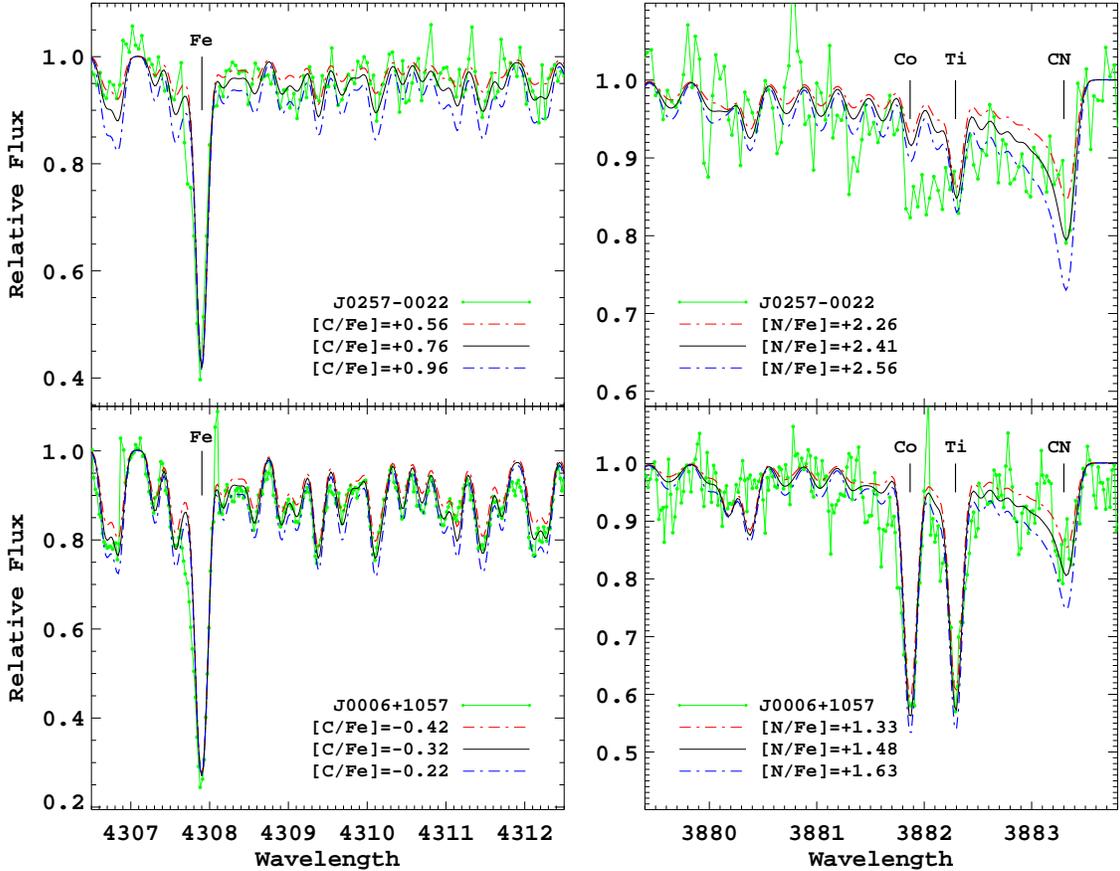}
\caption{Left: examples of fitting of the CH band used to determine the C abundance.
Right: examples of fitting of the CN band used to determine the N abundance.
The dotted line filled circle represents the observed spectra; the solid line refers to the best fit,
and the upper and lower dash-dotted lines correspond to synthetic spectra with changes of
1$\sigma$ fitting uncertainty in \AB{C}{Fe} and \AB{N}{Fe}.
The upper and lower panels refer to the fitting of the spectra of J0257$-$0022 and J0006+1057, respectively.
\label{fig:synfit_CN}}
\end{figure*}

\subsection{Carbon and Nitrogen}\label{subsec:CandN}

The carbon abundances of the program stars were derived by matching the
observed CH-AX band head at 4310\,{\AA} ~(i.e., the G band)
to the synthetic spectra. Examples of
such fits are presented in the left panels of Figure~\ref{fig:synfit_CN}.
The molecular line data for the spectrum synthesis were taken from the
Kurucz database. \footnote{http://kurucz.harvard.edu/molecules.html} 
The line broadening was determined using a single Gaussian profile
representing the combined effects from the instrumental profile, 
atmospheric turbulence, and stellar rotation. The width of the Gaussian
was determined by fitting a number of isolated and clear \ion{Fe}{1}
and \ion{Ti}{2} lines whose abundances have been derived from EW measurements.

The nitrogen abundance was measured from the CN-BX electronic
transition at 3883\,{\AA}, by comparing the observed spectra with synthetic spectra
generated with MOOG (see the right panels of Figure~\ref{fig:synfit_CN}
for examples). We could only determine the nitrogen abundances
of seven of the program stars whose CN-BX features are strong enough
for reliable measurements. We did not use the NH band at 3360\,{\AA},
because of an S/N (typically lower than five) of our spectra 
at that spectral region that is too low to perform spectral synthesis.

The carbon and nitrogen abundances of the program stars are listed in
Table~\ref{tab:abundance-part1}.

\subsection{Abundances Determined from Atomic Lines}

For 19 elements between Na and Ba, the abundances were computed using
the measured equivalent widths and the atomic data presented in
Table~\ref{tab:EW}, and the stellar parameters in Table~\ref{tab:stellar-param}.

We performed sanity checks for cases where the abundance of an element
was derived from a single line, or when abundances deviated by more than
3$\,\sigma$ from the average computed for an atomic species from
multiple lines. In these cases, the abundances were verified with
spectral synthesis and modified if necessary.
For clearly detected and isolated lines, synthetic spectra with
corresponding abundances derived from the measured equivalent widths
could well match the observed spectra.
However, for lines that
suffer from blending or problems in setting the continuum level,
the abundance derived from equivalent widths could not properly
match the observed spectral line, and hence the abundance derived
from spectral synthesis was adopted.
Table~\ref{tab:EW} lists the used atomic line data, the
measured equivalent widths, as well as the derived abundances, where
lines marked as ``syn'' refer to abundances determined with spectral
synthesis.

\begin{figure}
\hspace{-1cm}\epsscale{1.2}
\plotone{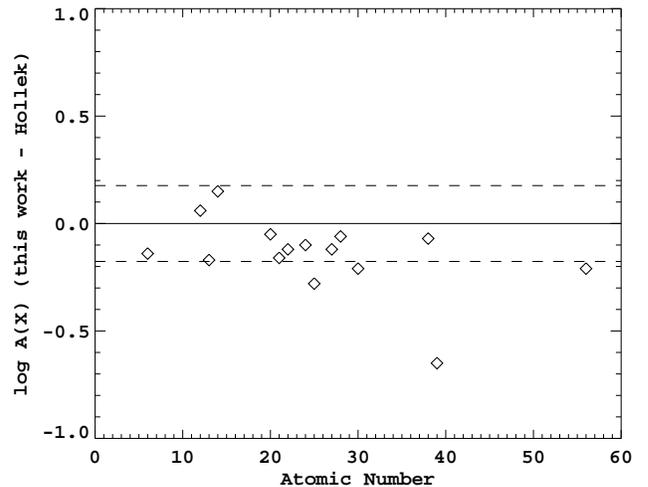}
\caption{Comparison of the abundances of J0326$+$0202 from this work
with the result from \citet{Hollek2011ApJ}, in the form of the abundance difference
vs. the atomic numbers. The dashed lines refer to the standard deviation of
the abundance difference ($\pm$0.18). The most deviated point corresponds to the element Y,
and is described in the text.
\label{fig:abun_compare}}
\end{figure}

For all of the program stars, the abundances of $\alpha$ elements (Mg, Si, Ca, and Ti)
have been determined. Among the light odd-Z elements (Na, Al, K, and Sc),
abundances of Sc have been determined for all of the program stars,
while for the warmer stars not all the lines are detectable,
e.g., the red doublet K lines are too weak in J0257$-$0022,
J1626$+$1721, and J1709$+$1616 to derive an abundance.
For Fe-peak elements (Cr, Mn, Co, Ni, and Zn), the abundances of Cr and Mn
could be determined for all of the program stars; meanwhile the abundances of Co, Ni,
and Zn could not be derived for the most metal-deficient object,
J1709$+$1616 due to its low Fe abundance and possibly insufficient S/N.
In the cases of the heavy elements (Sr, Y, Ba, and Eu),
the abundances of Sr and Ba were determined for all of the program stars
except J1709$+$1616 (the most metal-poor) and J1626$+$1721, respectively;
meanwhile abundances of Y could only be determined for four of these stars,
and the Eu lines have been detected and measured in only two objects,
J0102$+$0428, and J0343$-$0227. All of the elemental abundances of the programs stars
are shown in Table~\ref{tab:abundance-part1}.
We have also compared the abundances of J0326$+$0202
with those from \citet{Hollek2011ApJ}, as shown in Figure~\ref{fig:abun_compare}.
The difference between the abundance from this work and Hollek
is $-0.14\pm0.18$, indicating a rather good agreement.
The only species with significant difference is Y.
Since our measurements have used three Y lines resulting in a
standard deviation of $0.08$ in the derived abundance, 
further information should be required to explain the difference.

%
%
\begin{deluxetable*}{lrrcrcrrcrcrrcrcrrcrcc}
\centering
\tablenum{4}
\tablecolumns{22}
\tabletypesize{\scriptsize}
\tablewidth{0pc}
\tablecaption{Abundances of Individual Elements for the Program Stars.\label{tab:abundance-part1}}
\tablehead{
\colhead{}& \multicolumn{4}{c}{LAMOST J0006$+$1057}& \colhead{}&
            \multicolumn{4}{c}{LAMOST J0102$+$0428}& \colhead{}&
            \multicolumn{4}{c}{LAMOST J0126$+$0135}& \colhead{}&
            \multicolumn{4}{c}{LAMOST J0257$-$0022}& \colhead{}& \colhead{Sun}\\
\cline{2-5}\cline{7-10}\cline{12-15}\cline{17-20}\\
\colhead{}& \colhead{log\,$\epsilon$(X)}& \colhead{\AB{X}{Fe}}& \colhead{$\sigma$}& \colhead{$N$}& \colhead{}&
            \colhead{log\,$\epsilon$(X)}& \colhead{\AB{X}{Fe}}& \colhead{$\sigma$}& \colhead{$N$}& \colhead{}&
	    \colhead{log\,$\epsilon$(X)}& \colhead{\AB{X}{Fe}}& \colhead{$\sigma$}& \colhead{$N$}& \colhead{}&
	    \colhead{log\,$\epsilon$(X)}& \colhead{\AB{X}{Fe}}& \colhead{$\sigma$}& \colhead{$N$}& \colhead{}&
	    \colhead{log\,$\epsilon$(X)}}
\startdata
CH(C)      &   4.85&$-$0.32&0.10&  1&&   4.75&$-$0.93&0.15&  1&&   4.35&$-$0.51&0.15&  1&&   6.95&   0.76&0.20&  1&&8.43\\
CN(N)      &   6.05&   1.48&0.15&  1&&   6.45&   1.37&0.15&  1&&   5.70&   1.44&0.20&  1&&   8.00&   2.41&0.15&  1&&7.83\\
\ion{Na}{1}&   3.05&   0.07&0.09&  1&&   3.59&   0.10&0.13&  2&&   2.75&   0.08&0.15&  2&&   ... &   ... &... &...&&6.24\\
\ion{Mg}{1}&   5.01&   0.66&0.08&  3&&   5.24&   0.39&0.13&  7&&   4.45&   0.42&0.13&  5&&   5.45&   0.09&0.12&  5&&7.60\\
\ion{Al}{1}&   2.61&$-$0.58&0.05&  1&&   3.64&$-$0.06&0.11&  2&&   2.20&$-$0.68&0.11&  2&&   3.44&$-$0.77&0.03&  2&&6.45\\
\ion{Si}{1}&   4.72&   0.47&0.03&  1&&   5.14&   0.38&0.12&  2&&   4.47&   0.53&0.03&  1&&   5.80&   0.53&0.07&  1&&7.51\\
\ion{K}{1} &   2.51&   0.74&0.01&  2&&   2.86&   0.58&0.08&  1&&   2.01&   0.55&0.17&  1&&   ... &   ... &... &...&&5.03\\
\ion{Ca}{1}&   3.46&   0.38&0.07&  8&&   3.84&   0.25&0.09& 18&&   3.21&   0.44&0.09&  4&&   4.38&   0.28&0.11&  6&&6.34\\
\ion{Sc}{2}&$-$0.09&   0.02&0.07&  5&&   0.23&$-$0.17&0.11&  7&&$-$0.63&$-$0.22&0.12&  5&&   1.04&   0.13&0.05&  2&&3.15\\
\ion{Ti}{1}&   1.84&   0.15&0.11&  6&&   2.19&$-$0.01&0.12& 21&&   1.74&   0.36&0.14&  3&&   3.32&   0.61&0.19&  2&&4.95\\
\ion{Ti}{2}&   1.87&   0.18&0.12& 12&&   2.27&   0.07&0.14& 45&&   1.35&$-$0.03&0.18& 16&&   3.18&   0.47&0.10& 15&&4.95\\
\ion{V}{2} &   0.96&   0.29&0.04&  1&&   1.22&   0.04&0.12&  3&&   0.60&   0.24&0.04&  1&&   ... &   ... &... &...&&3.93\\
\ion{Cr}{1}&   1.91&$-$0.47&0.16&  5&&   2.65&$-$0.24&0.14& 15&&   1.43&$-$0.64&0.14&  6&&   3.18&$-$0.22&0.08&  3&&5.64\\
\ion{Cr}{2}&   ... &   ... &... &...&&   3.09&   0.20&0.07&  1&&   ... &   ... &... &...&&   ... &   ... &... &...&&5.64\\
\ion{Mn}{1}&   1.67&$-$0.50&0.11&  4&&   2.12&$-$0.56&0.12&  5&&   0.86&$-$1.00&0.06&  3&&   2.65&$-$0.54&0.09&  3&&5.43\\
\ion{Fe}{1}&   4.24&   0.00&0.14& 94&&   4.75&   0.00&0.15&164&&   3.93&   0.00&0.19& 61&&   5.26&   0.00&0.14& 60&&7.50\\
\ion{Fe}{2}&   4.24&   0.00&0.13& 11&&   4.76&   0.01&0.10& 19&&   3.93&   0.00&0.15&  4&&   5.25&$-$0.01&0.14&  3&&7.50\\
\ion{Co}{1}&   1.87&   0.14&0.04&  1&&   2.13&$-$0.11&0.16&  7&&   1.59&   0.17&0.01&  2&&   3.03&   0.28&0.12&  1&&4.99\\
\ion{Ni}{1}&   2.90&$-$0.06&0.06&  3&&   3.25&$-$0.22&0.18&  6&&   2.57&$-$0.08&0.07&  1&&   4.07&   0.09&0.12&  2&&6.22\\
\ion{Zn}{1}&   1.72&   0.42&0.04&  2&&   1.80&$-$0.01&0.10&  1&&   1.43&   0.44&0.07&  2&&   2.71&   0.39&0.14&  1&&4.56\\
\ion{Sr}{2}&$-$2.46&$-$2.07&0.06&  2&&$-$1.06&$-$1.18&0.14&  2&&$-$2.31&$-$1.61&0.04&  1&&   1.05&   0.42&0.12&  2&&2.87\\
\ion{Y}{2} &   ... &   ... &... &...&&$-$1.62&$-$1.08&0.03&  3&&   ... &   ... &... &...&&$-$0.08&$-$0.05&0.11&  1&&2.21\\
\ion{Ba}{2}&$-$2.83&$-$1.75&0.03&  1&&$-$1.26&$-$0.69&0.13&  5&&$-$2.53&$-$1.14&0.16&  3&&$-$0.17&$-$0.11&0.00&  2&&2.18\\
\ion{Eu}{2}&   ... &   ... &... &...&&$-$2.13&   0.10&0.08&  1&&   ... &   ... &... &...&&   ... &   ... &... &...&&0.52\\
\enddata
\end{deluxetable*}

%
%
\begin{deluxetable*}{lrrcrcrrcrcrrcrcrrcrcc}
\centering
\tablenum{4}
\tablecolumns{22}
\tabletypesize{\scriptsize}
\tablewidth{0pc}
\tablecaption{Continued.}
\tablehead{
\colhead{}& \multicolumn{4}{c}{LAMOST J0326$+$0202}& \colhead{}&
            \multicolumn{4}{c}{LAMOST J0343$-$0227}& \colhead{}&
            \multicolumn{4}{c}{LAMOST J1626$+$1721}& \colhead{}&
            \multicolumn{4}{c}{LAMOST J1709$+$1616}& \colhead{}& \colhead{Sun}\\
\cline{2-5}\cline{7-10}\cline{12-15}\cline{17-20}\\
\colhead{}& \colhead{log\,$\epsilon$(X)}& \colhead{\AB{X}{Fe}}& \colhead{$\sigma$}& \colhead{$N$}& \colhead{}&
            \colhead{log\,$\epsilon$(X)}& \colhead{\AB{X}{Fe}}& \colhead{$\sigma$}& \colhead{$N$}& \colhead{}&
	    \colhead{log\,$\epsilon$(X)}& \colhead{\AB{X}{Fe}}& \colhead{$\sigma$}& \colhead{$N$}& \colhead{}&
	    \colhead{log\,$\epsilon$(X)}& \colhead{\AB{X}{Fe}}& \colhead{$\sigma$}& \colhead{$N$}& \colhead{}&
	    \colhead{log\,$\epsilon$(X)}}
\startdata
CH(C)      &   5.15&   0.08&0.10&  1&&   5.90&$-$0.11&0.15&  1&&   6.30&   1.07&0.20&  1&&   6.30&   1.58&0.20&  1&&8.43\\
CN(N)      &   5.70&   1.23&0.15&  1&&   6.35&   0.94&0.15&  1&&   ... &   ... &... &...&&   ... &   ... &... &...&&7.83\\
\ion{Na}{1}&   3.46&   0.58&0.01&  2&&   3.26&$-$0.02&0.14&  2&&   3.02&$-$0.02&0.03&  2&&   ... &   ... &... &...&&6.24\\
\ion{Mg}{1}&   4.99&   0.75&0.17&  8&&   5.58&   0.40&0.03&  4&&   4.90&   0.50&0.16&  3&&   4.17&   0.28&0.12&  3&&7.60\\
\ion{Al}{1}&   2.36&$-$0.73&0.04&  1&&   3.50&$-$0.53&0.06&  1&&   2.92&$-$0.33&0.08&  1&&   ... &   ... &... &...&&6.45\\
\ion{Si}{1}&   4.93&   0.78&0.07&  2&&   5.52&   0.43&0.04&  1&&   4.77&   0.46&0.09&  1&&   3.98&   0.18&0.10&  1&&7.51\\
\ion{K}{1} &   2.49&   0.82&0.03&  2&&   3.12&   0.51&0.08&  1&&   ... &   ... &... &...&&   ... &   ... &... &...&&5.03\\
\ion{Ca}{1}&   3.40&   0.42&0.08& 12&&   4.23&   0.31&0.11& 14&&   3.41&   0.27&0.08&  2&&   3.05&   0.42&0.11&  1&&6.34\\
\ion{Sc}{2}&$-$0.23&$-$0.02&0.04&  5&&   0.65&$-$0.08&0.07&  5&&   0.02&   0.07&0.11&  1&&$-$0.37&   0.19&0.14&  1&&3.15\\
\ion{Ti}{1}&   1.81&   0.22&0.06&  7&&   2.70&   0.17&0.11& 15&&   ... &   ... &... &...&&   ... &   ... &... &...&&4.95\\
\ion{Ti}{2}&   1.75&   0.16&0.10& 32&&   2.79&   0.26&0.11& 27&&   1.92&   0.17&0.11& 11&&   1.90&   0.66&0.13& 10&&4.95\\
\ion{V}{2} &   0.60&   0.03&0.03&  1&&   1.52&   0.01&0.05&  1&&   ... &   ... &... &...&&   ... &   ... &... &...&&3.93\\
\ion{Cr}{1}&   1.92&$-$0.36&0.16&  6&&   3.00&$-$0.22&0.08&  9&&   2.19&$-$0.25&0.06&  3&&   1.64&$-$0.29&0.15&  1&&5.64\\
\ion{Cr}{2}&   ... &   ... &... &...&&   3.31&   0.09&0.06&  1&&   ... &   ... &... &...&&   ... &   ... &... &...&&5.64\\
\ion{Mn}{1}&   1.42&$-$0.65&0.09&  3&&   2.80&$-$0.21&0.09&  6&&   2.04&$-$0.19&0.10&  1&&   1.65&$-$0.07&0.11&  1&&5.43\\
\ion{Fe}{1}&   4.14&   0.00&0.10&127&&   5.08&   0.00&0.13&111&&   4.30&   0.00&0.13& 46&&   3.79&   0.00&0.14& 24&&7.50\\
\ion{Fe}{2}&   4.13&$-$0.01&0.11&  7&&   5.08&   0.00&0.15& 14&&   ... &   ... &... &...&&   ... &   ... &... &...&&7.50\\
\ion{Co}{1}&   1.82&   0.19&0.10&  5&&   2.60&   0.03&0.07&  2&&   2.42&   0.63&0.10&  1&&   ... &   ... &... &...&&4.99\\
\ion{Ni}{1}&   2.89&   0.03&0.09&  5&&   3.90&   0.10&0.09&  7&&   3.19&   0.17&0.14&  2&&   ... &   ... &... &...&&6.22\\
\ion{Zn}{1}&   1.42&   0.22&0.04&  2&&   2.35&   0.21&0.06&  1&&   ... &   ... &... &...&&   ... &   ... &... &...&&4.56\\
\ion{Sr}{2}&$-$1.18&$-$0.69&0.01&  2&&   0.29&$-$0.16&0.07&  2&&$-$0.84&$-$0.51&0.03&  2&&   ... &   ... &... &...&&2.87\\
\ion{Y}{2} &$-$1.98&$-$0.83&0.08&  3&&$-$0.58&$-$0.37&0.09&  4&&   ... &   ... &... &...&&   ... &   ... &... &...&&2.21\\
\ion{Ba}{2}&$-$2.55&$-$1.37&0.03&  2&&$-$0.58&$-$0.34&0.18&  4&&   ... &   ... &... &...&&$-$1.54&$-$0.01&0.18&  1&&2.18\\
\ion{Eu}{2}&   ... &   ... &... &...&&$-$1.99&$-$0.09&0.04&  2&&   ... &   ... &... &...&&   ... &   ... &... &...&&0.52\\
\enddata
\tablecomments{{\footnotesize The column of ``$N$'' refers to the number of lines
adopted for determination of the elemental abundances.
The photospheric solar abundances from \citet{Asplund2009ARAA} are also shown for reference.}}
\end{deluxetable*}

%
\subsection{Abundance Error Estimation}\label{subsec:abun_error}

The error of the derived abundances mainly arises from two aspects,
the uncertainties in the equivalent width measurements 
and those caused by the uncertainties from the stellar parameters.

When $N \ge 2$ lines of individual species of an element were observed, 
the dispersion in the measurements of multiple lines around
the average abundances, i.e., $\sigma$log\,$\epsilon$(X),
was used to present the error caused by random uncertainties
in the equivalent widths, as given in Table~\ref{tab:abundance-part1}.
If the elemental abundance was determined from a single line,
the statistic error of the equivalent widths was estimated
based on the classical formula of \citet{Cayrel1988IAUS}:
\begin{equation}
\sigma_{EW} = \langle \Delta W ^{2} \rangle ^{1/2} \simeq 1.6 (w \Delta x)^{1/2} \epsilon
\label{equa:error}
\end{equation}
where $w$ is the FWHM of the line, $\Delta x$ refers to the sampling step of the MIKE spectra,
and $\epsilon$ refers to the reciprocal spectral S/N in the case of normalized spectra.
The uncertainty in the derived abundance corresponding to 2$\sigma_{EW}$
was adopted as a conservative estimate of the random error caused by
the equivalent width measurement. The errors relevant to the uncertainties
of the equivalent width measurement are listed in Table~\ref{tab:EW}
for each individual line that has been used for abundance estimation.

The abundance uncertainties associated with the uncertainties of
the stellar parameters were estimated by varying {\Tefft} by $+$150\,K,
{\logg} by $+$0.3\,dex, and $\xi$ by $+$0.3\,km\,s$^{-1}$
in the stellar atmospheric model. Table~\ref{tab:abun-sigma}
summarizes the corresponding abundance uncertainties 
and the total uncertainty for the three errors, which was computed
as the quadratic sum, for J0326$+$0202 as an example.
We noted that the abundance uncertainties caused by
the equivalent width measurements are generally smaller than
those propagating from the stellar parameters, and therefore,
we have adopted the latter uncertainties as reference error bars
for further discussions concerning the abundances of the sample.

%
\begin{figure*}
\epsscale{0.85}
\plotone{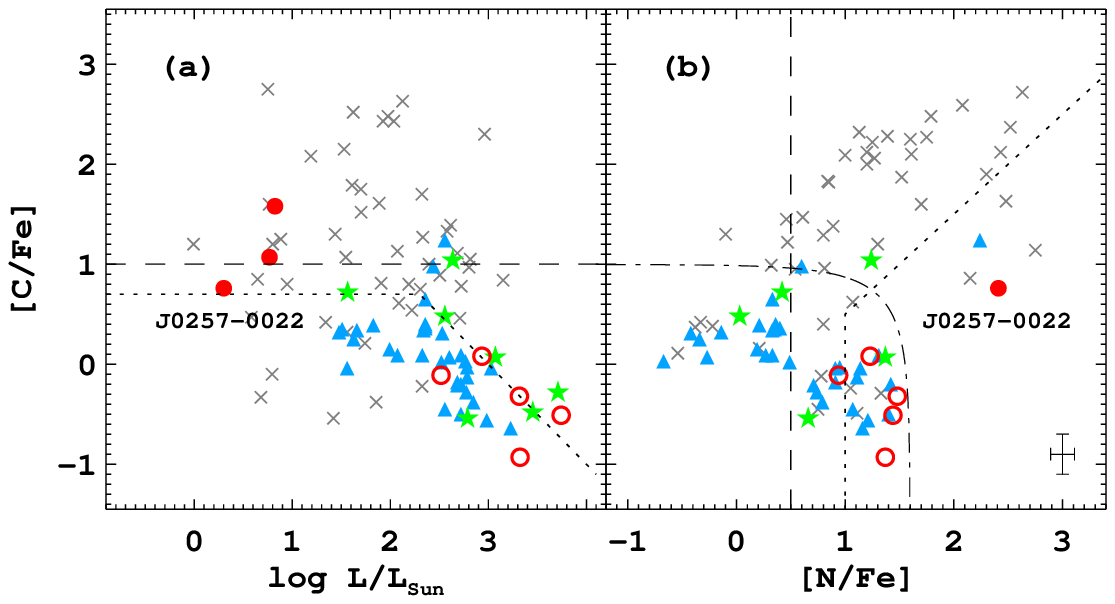}
\caption{
(a) \AB{C}{Fe} ratio as a function of the luminosity estimated from {\Tefft} and {\logg}
of our sample. The dotted line indicates the dividing line between carbon-enhanced
and carbon-normal stars as defined in \citet{Aoki2007ApJ}. The dashed line corresponds to \ABeq{C}{Fe}{1.0}.
(b) \AB{C}{Fe} vs. \AB{N}{Fe}. The two criteria for NEMP stars suggested by \citet{Pols2012AA}
are respectively shown in dashed (\ABge{N}{Fe}{1.0} and \ABge{N}{C}{0.5}) 
and dash-dotted lines (\ABgt{(C+N)}{Fe}{0.9}).
The filled and open circles respectively refer to carbon-enhanced and carbon-normal stars.
Metal-poor stars from ``First Stars'' \citep{Spite2005AA}, \citet{Yong2013ApJa}, and \citet{Placco2014ApJ}
are also plotted for comparison, in filled triangles, crosses, and filled pentacles, respectively.
The potential CNEMP object J0257$-$0022 is marked.
\label{fig:CandN}}
\end{figure*}

\section{RESULTS AND INTERPRETATION}\label{sec:results}

\subsection{CEMP and NEMP Objects}\label{subsec:CEMPandNEMP}

%
%
\begin{deluxetable}{lrrrr}
\tablenum{5}
\tablecolumns{5}
\tabletypesize{\scriptsize}
\tablewidth{0pc}
\tablecaption{Uncertainties of log\,$\epsilon$(X) Propagated from the Stellar Parameters
(as Described in Section~\ref{subsec:abun_error}),
Computed for J0326$+$0202 as an Example.\label{tab:abun-sigma}}
\tablehead{
\colhead{Ion}& \colhead{$\Delta$\Tefft}& \colhead{$\Delta$log $g$}& \colhead{$\Delta \xi$}& \colhead{$\sigma_{\mbox{tot}}$}\\
\colhead{}& \colhead{$+$ 150\,K}& \colhead{$+$ 0.3\,dex}& \colhead{$+$ 0.3\,km s$^{-1}$}& \colhead{}}
\startdata
CH(C)       &0.30&$-$0.15&$-$0.05&0.34\\
CN(N)       &0.25&$-$0.05&   0.00&0.25\\
\ion{Na}{1} &0.17&$-$0.03&$-$0.18&0.25\\
\ion{Mg}{1} &0.13&$-$0.06&$-$0.08&0.16\\
\ion{Al}{1} &0.15&$-$0.05&$-$0.18&0.24\\
\ion{Si}{1} &0.16&$-$0.02&$-$0.03&0.16\\
\ion{K}{1}  &0.12&$-$0.02&$-$0.02&0.12\\
\ion{Ca}{1} &0.11&$-$0.03&$-$0.04&0.12\\
\ion{Sc}{2} &0.10&   0.08&$-$0.08&0.15\\
\ion{Ti}{1} &0.19&$-$0.03&$-$0.02&0.19\\
\ion{Ti}{2} &0.08&   0.08&$-$0.08&0.14\\
\ion{V}{2}  &0.08&   0.09&$-$0.01&0.12\\
\ion{Cr}{2} &0.18&$-$0.04&$-$0.09&0.21\\
\ion{Mn}{1} &0.21&$-$0.05&$-$0.14&0.26\\
\ion{Fe}{1} &0.17&$-$0.04&$-$0.10&0.20\\
\ion{Fe}{2} &0.02&   0.09&$-$0.06&0.11\\
\ion{Co}{1} &0.21&$-$0.03&$-$0.07&0.22\\
\ion{Ni}{1} &0.20&$-$0.05&$-$0.20&0.29\\
\ion{Zn}{1} &0.09&   0.04&   0.01&0.10\\
\ion{Sr}{2} &0.11&   0.07&$-$0.27&0.30\\
\ion{Y}{2}  &0.13&   0.08&$-$0.01&0.15\\
\ion{Ba}{2} &0.14&   0.08&$-$0.03&0.16\\
\enddata
\end{deluxetable}

Despite the limited sample size, there are several objects
in our sample exhibiting relatively high \AB{C}{Fe} ratios,
as shown in Figure~\ref{fig:CandN}. To check the fraction of CEMP stars, 
we have adopted the classification of \citet{Aoki2007ApJ}, who suggest a scheme that takes
into consideration the nucleosynthesis and mixing effects in
giants. This is important for our sample, because five of our nine stars
are giants (see Figure~\ref{fig:HR}). We therefore consider a star to be
CEMP when \ABge{C}{Fe}{+0.7} in the case of stars with luminosities of
$\log(L/L_{\odot}) \le 2.3$, or when \ABge{C}{Fe}{+3.0}$-\log(L/L_{\odot})$
in the case of stars with $\log(L/L_{\odot}) > 2.3$. The luminosities of our sample stars were
computed based on the prescription of \citet{Aoki2007ApJ}, assuming a
typical mass of $M=0.8M_{\odot}$ for halo stars.

Figure~\ref{fig:CandN}a shows that there are four objects in our sample
that are above the limit, among which there is a red giant J0126$+$0135
that is located slightly off the limit and has very low \ABeq{C}{Fe}{-0.51}).
Therefore, we have decided to classify only the three objects with \ABgt{C}{Fe}{0.7}
as CEMP stars (filled circles in Figure~\ref{fig:CandN}) 
and the rest as carbon-normal objects (open circles).

It is known that in addition to \AB{C}{Fe}, \AB{N}{Fe} is notably enhanced in
many CEMP stars as well, and although this phenomenon appears to be
rare, a population of nitrogen-enhanced metal-poor (NEMP) stars with
\ABgt{N}{C}{0.5} has been discovered \citep{Johnson2007ApJ,Pols2012AA}.
We have adopted the criteria of \citet{Pols2012AA} for classifying stars as
NEMP, i.e., \ABge{N}{Fe}{1.0} and \ABge{N}{C}{0.5}.
In Figure~\ref{fig:CandN}(b), it can be seen that among the six program stars
with determined nitrogen abundances, there are five that are on the right side 
of the dotted line, and hence are nitrogen-enhanced. However, 
as can be seen in Figure~\ref{fig:HR} and of Figure~\ref{fig:CandN}(a), 
four of them have \ABlesssim{C}{Fe}{0.0} and \AB{N}{Fe} between 1.2 through 1.5,
and are on the upper part of the red giant branch, and thus they may not be 
genuine NEMP, as will be discussed in Section~\ref{subsec:mixing}.

If we use the separation of so-called ``mixed'' from ``unmixed'' stars
defined as in \citet{Spite2005AA} (i.e., the dashed line in Figure~\ref{fig:CandN}(b),
corresponding to \ABeq{N}{Fe}{0.5}), the four
objects belong to the group of ``mixed'' stars. In these stars,
nitrogen has been produced from the burning of carbon, and hence they are
usually carbon-poor and nitrogen-rich. The same holds true when the
alternative criterion of \citet{Pols2012AA} of \ABgt{(C+N)}{Fe}{0.9} is
considered (the dash-dotted line in Figure~\ref{fig:CandN}(b)).
The CEMP turn-off star J0257$-$0022 is well located in the NEMP region,
which makes it a potential carbon and nitrogen-enhanced metalpoor (CNEMP) star \citep{Pols2012AA}.
We will further discuss this point in Section~\ref{subsec:mixing}.

\subsection{Abundance Trends of the Light Elements}\label{subsec:light-el}

\begin{figure*}
\epsscale{1.10}
\plotone{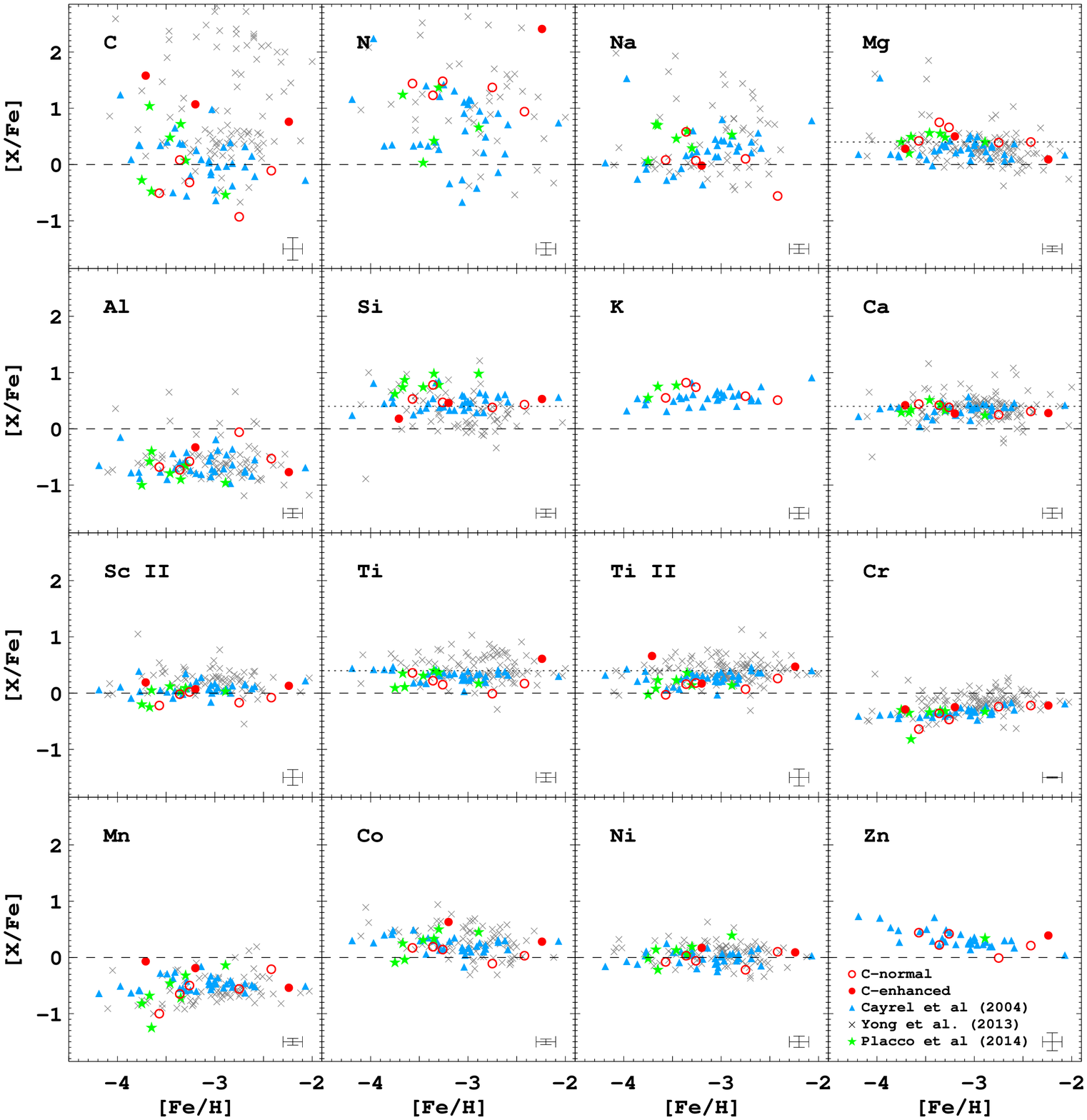}
\caption{\AB{X}{Fe} vs. \FeH ~of 15 elements (16 species) lighter than Zn.
For all $\alpha$ elements (Mg, Si, Ca, and Ti), the canonical value of \ABsim{$\alpha$}{Fe}{+0.4}
for the halo stars \citep{McWilliam1997ARAA} is plotted for reference.
The meanings of the different symbols are the same as in Figure~\ref{fig:CandN}.
\label{fig:abun_C2Zn}}
\end{figure*}

The abundance ratios \AB{X}{Fe} of the program stars are plotted against
{\FeH} in the right panels of Figure~\ref{fig:abun_C2Zn} 
for the elements from C through Zn. The
observed chemical abundances of metal-poor stars from the ``First
Stars'' program \citep{Cayrel2004AA}, and from
\citet{Yong2013ApJa} and \citet{Placco2014ApJ}, are also plotted
for comparison. Note that the ``First Stars'' sample has been
re-analyzed by \citet{Yong2013ApJa}, but since there are a few elements
not included in their analysis, we have adopted the abundances of
the sample of ``First Stars'' from \citet{Cayrel2004AA},
\citet{Spite2005AA}, and \citet{Francois2007AA}.

%
%

In general, the abundance ratios seen in our sample agree well with the
abundance ratio trends defined by the literature samples. However,
for the most iron-deficient object in our sample (J1709$+$1616), and
the carbon and nitrogen-enhanced object J0257$-$0022, a few elements (in
particular, Ti, Mn, and Ba) show slight deviations from these trends.

Compared with those of other elements, dispersions in the distributions
of C, N, Na, and Al are notably larger.
Unlike the behavior found by \citet{Yong2013ApJa},
the abundance ratios of the $\alpha$ elements Mg and Si (as well as Ca)
of the program stars distribute around the typical halo value of
\ABeq{$\alpha$}{Fe}{0.4} \citep[e.g., ][the dotted lines in Figure~\ref{fig:abun_C2Zn}]
{McWilliam1997ARAA}.

Among the iron-peak elements, a small scatter and clear dependence on Fe
are found for Cr and Zn. The abundance ratio of \AB{Cr}{Fe} exhibits the
lowest observed scatter together with a positive slope, which agrees well
with the results of \citet{Cayrel2004AA}. These authors interpreted
this trend as an indication that Cr and Fe are produced in the same
nucleosynthesis process, so that mixing in the interstellar medium would
not result in any significant scatter around the abundance trend.
Furthermore, a clear anti-correlation between \AB{Zn}{Fe} and {\FeH} is
seen. This suggests that neutron-capture processes may not have
contributed much to the production of Zn in progenitors of these EMP
stars, since if that is the case, an anti-correlation versus metallicities
would be expected for the abundance ratio.

\subsection{Neutron-capture Elements}\label{subsec:heavy-el}

\begin{figure*}
\epsscale{0.65}
\plotone{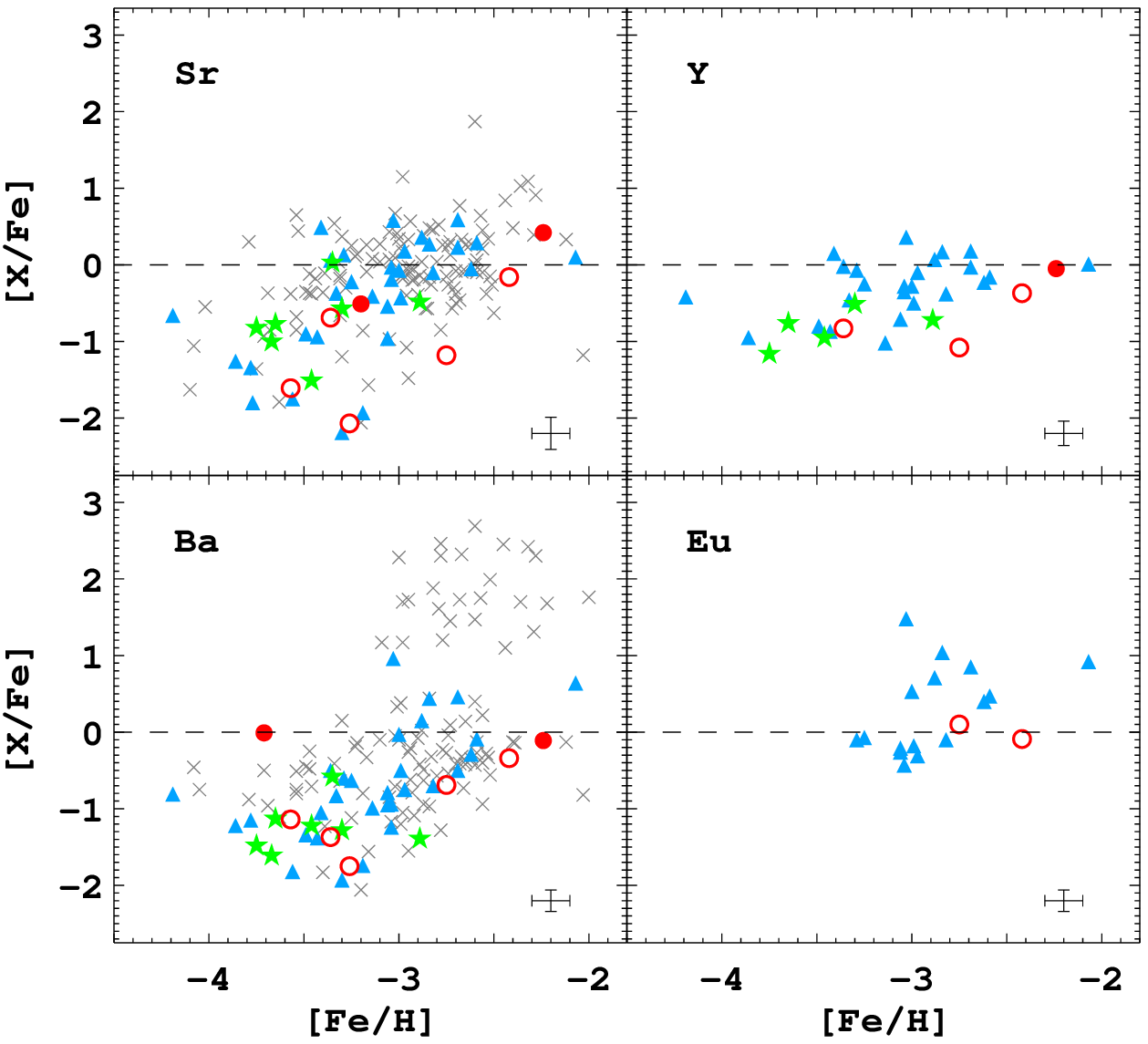}
\caption{\AB{Sr}{Fe}, \AB{Y}{Fe}, \AB{Ba}{Fe} and \AB{Eu}{Fe} vs. {\FeH} of the program stars.
The meanings of the different symbols are the same as in Figure~\ref{fig:CandN},
except that the triangles refer to data from \citet{Francois2007AA}.
\label{fig:abun_heavy}}
\end{figure*}

The abundance ratios of the neutron-capture elements Sr, Y, Ba, and Eu
versus {\FeH} are shown in Figure~\ref{fig:abun_heavy}.
In agreement with previous investigations, very large dispersions are seen,
especially at \FeHlt{-2.8} \citep{Ryan1996ApJ,Cayrel2004AA}. For all of 
the program stars with measurable Ba abundances, the \AB{Ba}{Fe} ratios
are not higher than the Solar value, which means that these two carbon-enhanced
stars (J0257$-$0022 and J1709+1616) belong to the class of CEMP-no stars.
Eu abundances are only measured in the two giants with relatively higher metallicities. 
Both giants show a near-Solar abundance ratio of \AB{Eu}{Fe}, 
and combined with the ``First Stars'' sample support the indication 
that metal-poor stars with high \AB{Eu}{Fe} are rare \citep{Francois2007AA}.

%
%

\begin{figure*}
\epsscale{0.8}
\plotone{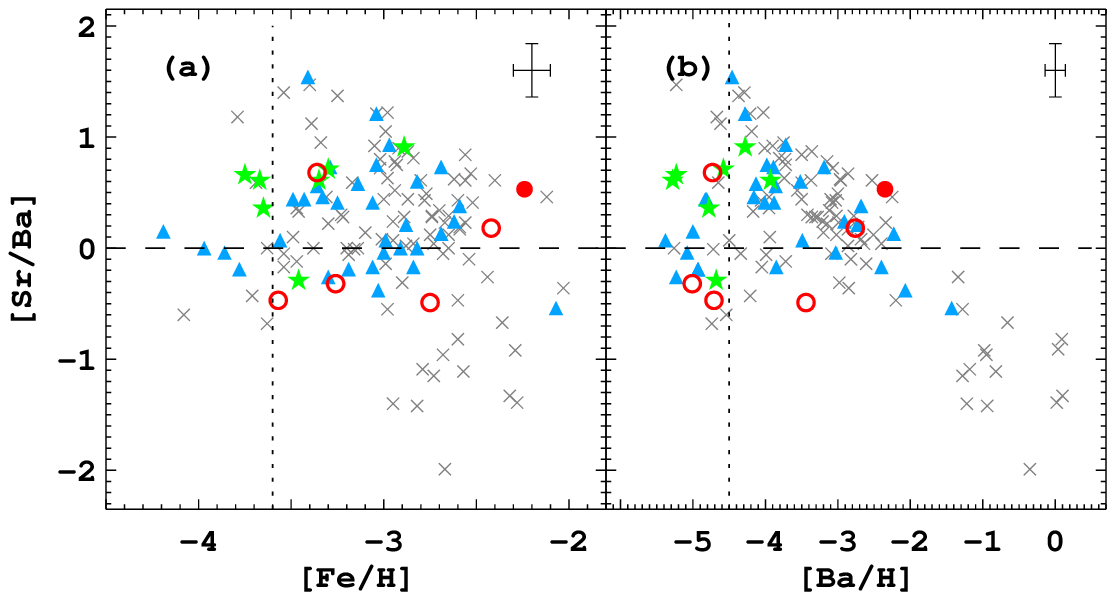}
\caption{\AB{Sr}{Ba} vs. {\FeH} (a) and \AB{Ba}{H} (b) of the program stars.
The meaning of different symbols are the same as in Figure~\ref{fig:abun_heavy}.
The horizontal dash-dotted lines refer to \ABsim{Sr}{Ba}{\pm0.4}, as guidance of
the abundance range to identify the weak and main $r$-processes.
The vertical dotted lines correspond to \FeHeq{-3.6} and \ABeq{Ba}{H}{-4.5}
in the two panels, respectively.
\label{fig:abun_SrBa}}
\end{figure*}

The abundance ratios of neutron-capture elements such as \AB{Sr}{Ba} can
be used to investigate the possible nucleosynthesis processes that have occurred
in the progenitors of EMP stars. \citet{Aoki2013ApJ} has found
several remarkable features in the distribution of \AB{Sr}{Ba} versus
{\FeH}, including a cutoff at approximately \FeHsim{-3.6}, and a
lower bound at about \ABsim{Sr}{Ba}{-0.5}.
Figure~\ref{fig:abun_SrBa}(a) shows the distribution of \AB{Sr}{Ba} for our
program stars and the metal-poor star samples from the literature 
as a function of {\FeH}. As can been seen, the number of observed
objects is indeed very limited around \ABsim{Sr}{Ba}{-0.5} and below.
However, from the currently available data set of metal-poor stars, it
does not seem clear whether the cutoff at \ABlt{Fe}{H}{-3.6} actually
exists. It is possible that there may be a cutoff or sharp drop
of the distribution of objects with \AB{Sr}{Ba} larger than the Solar
value at even lower metallicities, but this would require a larger body of
data for EMP and UMP stars for further exploration.

%
%

Figure~\ref{fig:abun_SrBa}(b) shows that if we adopt Ba as
a reference element, a major trend of anti-correlation between \AB{Sr}{Ba}
and \AB{Ba}{H} can be found all the way down to \ABsim{Ba}{H}{-4.5},
while the abundance ratio reverts back to a nearly Solar value
when it approaches to lower \AB{Ba}{H}. This is quite agreeable with
the results of \citet{Francois2007AA}, and we suspect that the
anti-correlation may extend to \AB{Ba}{H} higher than $-1.5$, e.g., to
around \ABsim{Ba}{H}{0.0}. The observed distribution trend of
\AB{Sr}{Ba} against \AB{Ba}{H} indicates that besides the main rapid
neutron-capture process, it is likely that an additional nucleosynthesis
mechanism such as a second neutron-capture process has contributed to
the observed amounts of Sr (and other neutron-capture elements) at
\ABgt{Ba}{H}{-4.5} \citep[e.g., ][]{Wanajo2001ApJ,Travaglio2004ApJ,Qian&Wasserburg2007PhR}.

\subsection{Abundance Patterns and Chemical Peculiarities}\label{subsec:pattern}

\begin{figure*}
\epsscale{1.1}
\plotone{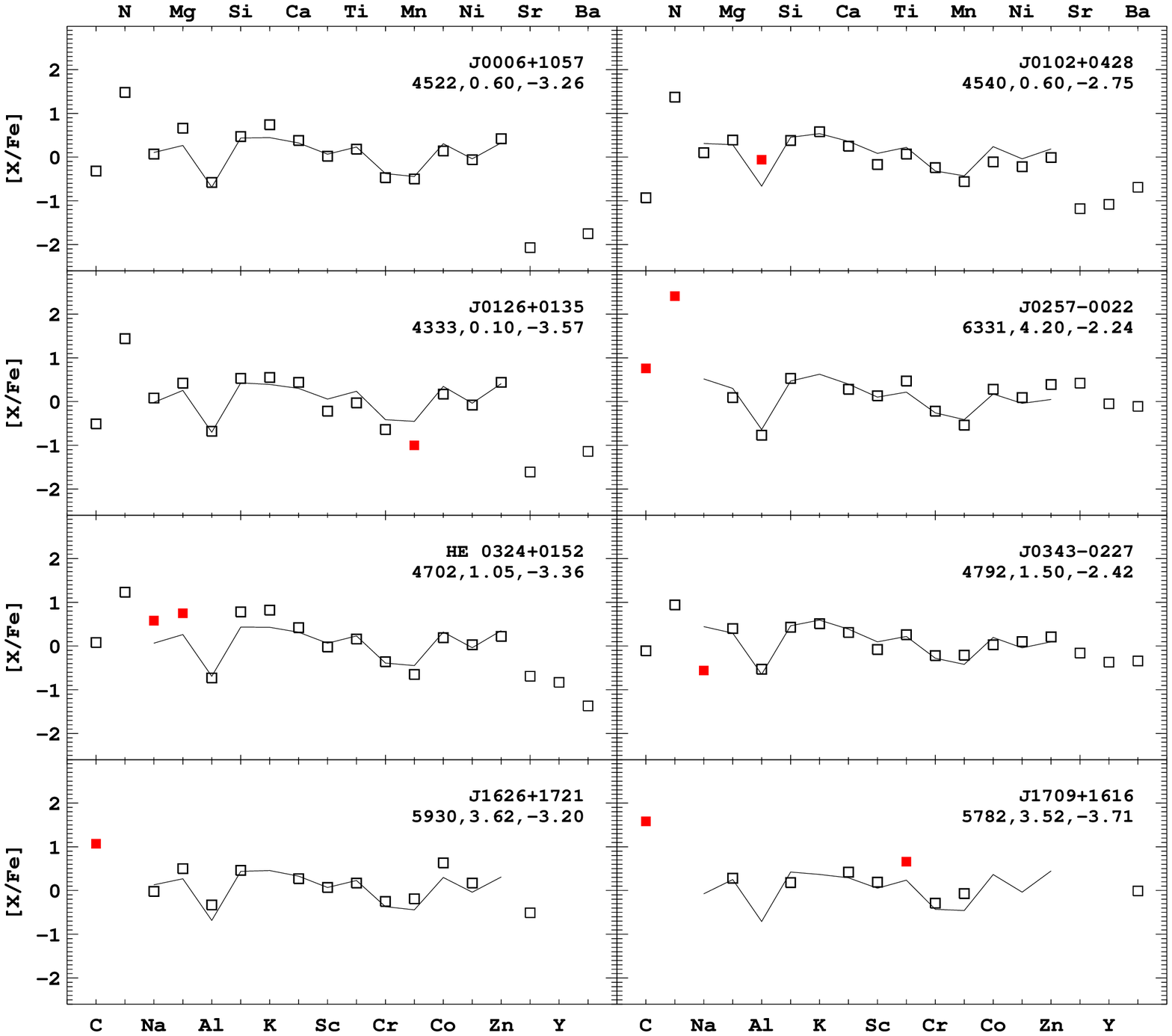}
\caption{Abundance pattern of the program stars. The solid line represents the ``average''
abundance derived from the regression coefficients from \citet{Cayrel2004AA}.
Filled squares refer to enhancement in carbon and/or nitrogen as defined in Section~\ref{subsec:CEMPandNEMP},
and elements with an abundance that differs from the solid line by more than 0.5\,dex.
The derived \Tefft, \logg, and \FeH ~together with the ID of each object are listed in the plot. 
\label{fig:abun_pattern}}
\end{figure*}

Despite the observed diversity in chemistry among low-metallicity stars,
based on homogeneously analyzed samples of metal-poor stars \citep{Cayrel2004AA,Yong2013ApJa}
it is believed that there exists a population that conforms to
the average abundance pattern, and thus is regarded as the chemically ``normal'' population.
Considering the fact that most of the carbon-normal objects in our sample are giants,
we have adopted the coefficients of the regression fit of \citet{Cayrel2004AA},
which were determined by means of a sample of metal-poor giants for which 
very high-quality spectra were obtained. The results of our comparison of the
abundance patterns are shown in Figure~\ref{fig:abun_pattern}.
For an element between Na and Zn, if the \AB{X}{Fe} ratio deviates more than 0.5\,dex
from the reference abundance pattern, it is regarded as abnormal and marked by filled squares.
For five out of the eight program stars, there is at least one element
showing such an abnormal abundance. However, if only carbon-normal objects with more than
one abnormal element are counted, only J0326$+$0202
exhibits a moderate over-abundance in Na and Mg.

%
%

Regarding neutron-capture elements, none of our stars are strongly
enhanced in Sr, Y, Ba, or Eu relative to the Solar abundance ratios.
However, we note that two of our stars, J0006$+$1057 and J0126$+$0135,
have quite low Sr and Ba abundances, respectively
with \AB{Sr}{H} of $-$5.33 and $-$5.18, and \AB{Ba}{H} of $-$5.01 and $-$4.71.
Using the stellar abundance data of over 1,000 stars
from the Galactic field and dwarf galaxies, \citet{Roederer2013AJ}
has noted that strontium and barium have been detected in all environments.
When comparing J0006$+$1057 and J0126$+$0135 with \citet{Roederer2013AJ}'s Figure 1
which presents the distribution of \AB{Ba}{H} versus \AB{Sr}{H}
in field stars and dwarf galaxies,
both of our program stars are located in the lower left corner of the plot,
and close to the detection thresholds. As pointed out by \citet{Roederer2013AJ},
such objects with unusually low \AB{Sr}{H} and \AB{Ba}{H} are of great interest,
in the sense that they raise the prospect that at the early phase of Galactic chemical evolution 
there is at least one kind of neutron-capture processes operating
as frequently as the nucleosynthesis mechanisms that
produce lighter elements such as $\alpha$ and iron-peak elements.
Interestingly, J0006-1057 and J0126+0135 have [Sr/Ba] ratios of $-0.3$ and
$-0.5$, respectively (see Table~\ref{tab:abundance-part1} and the right panel of Figure~\ref{fig:abun_SrBa}).
These values very much resemble those seen in strongly $r$-process-enhanced stars,
such as CS22892-052 (\ABeq{Sr}{Ba}{-0.4}; \citealt{Sneden2003ApJ}),
rather than the values typical for stars showing the abundance
signature of the weak $r$-process, e.g., HD122563 (\ABeq{Sr}{Ba}{+0.8};
\citealt{Honda2006ApJ}). This suggests that not only the
weak $r$-process, but also the main $r$-process contributed significantly to
the inventory of neutron-capture elements during the earliest phases of
Galactic chemical evolution.

Given the fact that \citet{Yong2013ApJa} found only seven
chemically peculiar objects in a sample of more than 100 carbon-normal
stars, the abundance patterns of our sample confirm that a chemically
``normal'' population does exist among metal-poor stars which may dominate, 
the low-metallicity region even at \FeHlt{-3.7}.

\subsection{``Mixed'' or ``Unmixed''?}\label{subsec:mixing}

\citet{Spite2005AA} observed a separation at \ABsim{C}{N}{-0.6}
among 35 EMP giants, and adopted such a separation
to divide their sample into ``mixed'' and ``unmixed'' groups.
Because the CNO process turns carbon into nitrogen,
the \AB{C}{N} ratio is believed to be a sensitive indicator of mixing; 
the observed separation demonstrates the signature of CN-cycle processing
in the surface abundances of these halo giants.
We thus compare the \AB{C}{N} abundance ratio
of our stars for objects with both measurements of carbon and nitrogen
in the samples of \citet{Spite2005AA}, \citet{Yong2013ApJa}, and \citet{Placco2014ApJ}.

%
%
\begin{figure*}
\epsscale{1.1}
\plotone{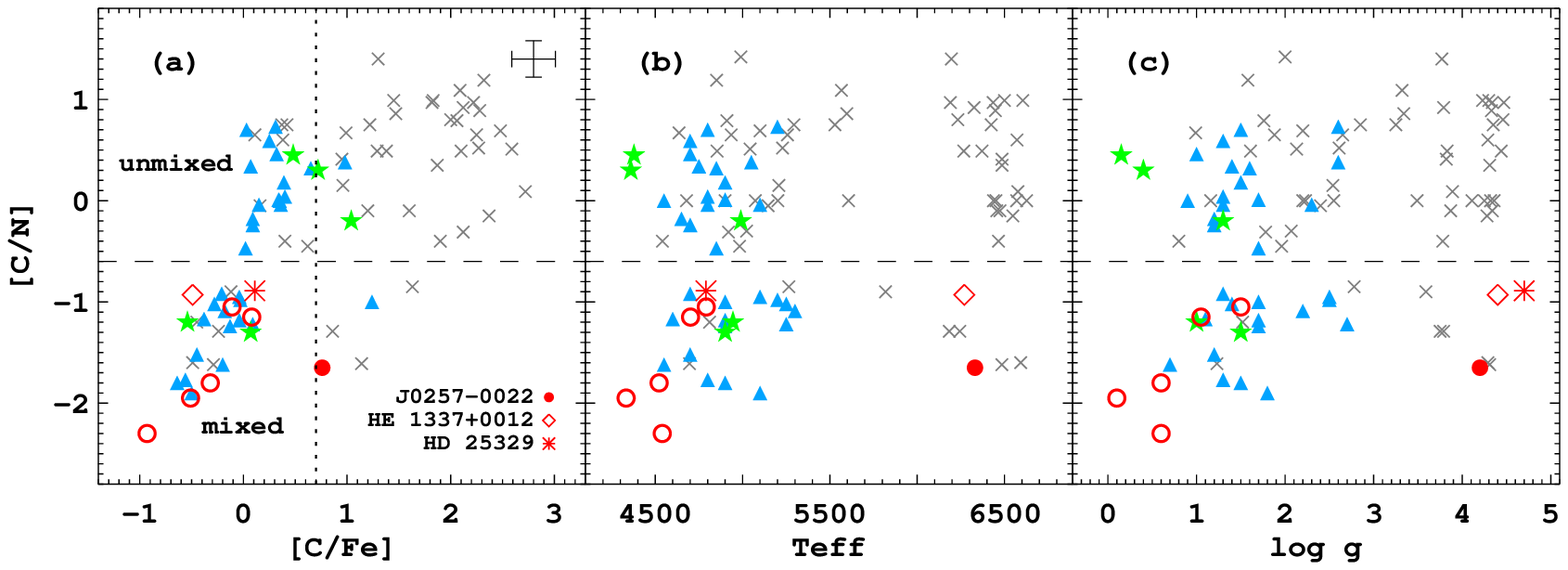}
\caption{
(a) \AB{C}{N} vs. \AB{C}{Fe},
(b) \AB{C}{N} vs. {\Tefft}, and
(c) \AB{C}{N} vs. {\logg} for the sample.
The plots only include objects with both measured abundances of carbon and nitrogen.
The dashed lines in all plots refer to \ABeq{C}{N}{-0.6} which is used by \citet{Spite2005AA}
to separate the mixed and unmixed stars.
The dotted lines in frame (a) refer to the adopted division of carbon-normal
and carbon-rich stars, i.e., \ABeq{C}{Fe}{0.7}, as described in the text.
The potential CNEMP object J0257$-$0022 (filled circle) is marked,
together with two other unevolved NEMP candidates, HE~1337+0012 (diamond)
and HD~25329 (star) from \citet{Pols2012AA}.
The meanings of the other symbols are the same as in Figure~\ref{fig:CandN}.
\label{fig:CN_CFe_Teff_logg}}
\end{figure*}

As shown in Figure~\ref{fig:CN_CFe_Teff_logg}a,
for the carbon-normal stars, i.e., the majority of the sample
from \citet[][filled triangles]{Spite2005AA}, \citet[][filled stars]{Placco2014ApJ},
and our program stars (open circles),
there are two distinct groups of stars
that can be separated by the dividing line at \ABsim{C}{N}{-0.6}.
We also note that there are very few carbon-enhanced objects
(the majority of the \citealt{Yong2013ApJa} sample and J0257$-$0022)
below the dividing line. This could be due to the fact that a great many 
of them are significantly enhanced in carbon,
and/or are main-sequence turnoff stars or dwarfs (as can be seen in 
Figure~\ref{fig:CN_CFe_Teff_logg}(b) and Figure~\ref{fig:CN_CFe_Teff_logg}(c))
whose deeper layers have not yet been dredged-up.

However, we note that the potential CNEMP turn-off star of our sample,
J0257$-$0022, lies well below the dividing line along with the ``mixed'' group.
This is rather interesting, since for this object the material processed
in the CN cycle has presumably not yet been dredged up from its interior to the outer layer.
For comparison, two unevolved NEMP candidates from \citet{Pols2012AA},
HE~1337$+$0012, and HD~25329 are also plotted in Figure~\ref{fig:CN_CFe_Teff_logg}.
Despite the different carbon abundances of these objects,
they are in a quite similar situation to J0257$-$0022,
e.g., unevolved (dwarf or main-sequence turnoff), nitrogen-enhanced,
and located in/close to the region of the ``mixed'' group.
As indicated in \citet{Pols2012AA}, ``genuine'' NEMP stars
are supposed to show signatures of accretion of material
that has undergone hot bottom burning (HBB), e.g.,
large enhancement in sodium and magnesium, while heavy $s$-process elements
such as barium and lead should not be significantly enhanced.
The low sodium abundance (\ABeq{Na}{Fe}{-1.13}, \citealt{Aoki2006ApJ})
of HE~1337$+$0012 basically rules out the possibility of accretion from HBB.
For HD~25329, only slight enhancements in sodium and magnesium are shown
(\ABeq{Na}{Fe}{+0.24}, and \ABeq{Mg}{Fe}{+0.59}, \citealt{Gratton2003AA}),
which does not support the material being accreted from an HBB process either.
For J0257$-$0022, a lack of enhancement in magnesium (\ABeq{Mg}{Fe}{+0.09})
and no measurable sodium lines make it very unlikely to be enriched 
by the accreted material from the HBB. However, the origin of the nitrogen-enhanced
status of these unevolved stars will remain unclear until more detailed
abundance analysis is available.

\section{CONCLUSIONS}\label{sec:conclusions}

Eight potential metal-poor stars have been selected from DR1 of the low-resolution
spectroscopic survey of LAMOST and follow-up observations in high-resolution spectroscopy
with Magellan/MIKE.
Based on the high-resolution analysis, we have confirmed five EMP stars,
including two with \FeHlt{-3.5}.
Among these objects, three are newly discovered, and one (J0126$+$0135)
is confirmed and analyzed using high-resolution spectra for the first time.

The abundances of C, N, 15 elements from Na to Zn, and 4 neutron-capture
elements Sr, Y, Ba, and Eu, have been derived for the sample.
When compared with results from the literature, no significant differences have been found.
Three out of the eight program stars are regarded as CEMP stars,
and one star with enhancement in both carbon and nitrogen has been discovered
and classified as a potential CNEMP star.

Efforts have been made to investigate the enrichment of the neutron-capture elements
of the EMP stars, especially the distribution of the abundance ratio \AB{Sr}{Ba},
as well as the comparison between the so-called ``mixed'' and ``unmixed'' groups of stars.
Our analysis indicates that enlarging the data set of objects with \FeHlt{-3.5}
is fairly important to fully understand the nucleosynthesis process of the progenitors
of the low-metallicity stars, e.g., the behavior of abundance ratios of neutron-capture elements
as well as to further explore the shape and the truth of a cutoff of the low-metallicity tail
of the MDF of the Galactic halo. The on-going LAMOST spectroscopic survey
will be able to improve the situation, and shed light on the nature of Galactic chemical evolution.

\acknowledgments
We are grateful to the anonymous referee who made valuable suggestions that
help improve the paper. 
H.N.L., G.Z., and W.W. acknowledge support by NSFC grants No. 11103030, 11233004, and 11390371.
L.W. is supported by the Young Researcher Grant of National Astronomical Observatories,
Chinese Academy of Sciences. N.C. acknowledges support from Sonderforschungsbereich 881
``The Milky Way System'' (subproject A4) of the German Research Foundation (DFG).
Guoshoujing Telescope (the Large Sky Area Multi-Object Fiber Spectroscopic Telescope, LAMOST)
is a National Major Scientific Project built by the Chinese Academy of Sciences.
Funding for the project has been provided by the National Development and Reform Commission.
LAMOST is operated and managed by the National Astronomical Observatories, Chinese Academy of Sciences.
This research uses data obtained through the Telescope Access Program (TAP), 
which has been funded by the Strategic Priority Research Program 
``The Emergence of Cosmological Structures'' (Grant No. XDB09000000), National Astronomical Observatories, 
Chinese Academy of Sciences, and the Special Fund for Astronomy from the Ministry of Finance.

\bibliographystyle{apj} 
\bibliography{mpstar,lamost} 

\end{document}